\documentclass[12pt]{article}
\usepackage{amsmath,amssymb,color}
\usepackage{cite,epsf}

\usepackage{graphicx,array} 
\newcommand{\be}{\begin{equation}}
\newcommand{\ee}{\end{equation}}
\newcommand{\beq}{\begin{equation}}
\newcommand{\eeq}{\end{equation}}
\newcommand{\bea}{\begin{eqnarray}}
\newcommand{\eea}{\end{eqnarray}}

\newcommand{\ba}{\begin{eqnarray}}
\newcommand{\ea}{\end{eqnarray}}
\newcommand{\s}{\sigma}

\begin{document}

\begin{titlepage}
\vspace{10pt}
\hfill
{\large\bf HU-EP-09/11}
\vspace{20mm} 
\begin{center}

{\Large\bf  On spacelike and timelike minimal surfaces in $AdS_n$}

\vspace{45pt}

{\large Harald Dorn${}^a$,
 George Jorjadze${}^{a,\,b}$ and Sebastian Wuttke${}^a$
{\footnote{dorn@physik.hu-berlin.de, jorj@physik.hu-berlin.de,
  wuttke@mathematik.hu-berlin.de }}}
\\[15mm]
{\it\ ${}^a$Institut f\"ur Physik der
Humboldt-Universit\"at zu Berlin,}\\
{\it Newtonstra{\ss}e 15, D-12489 Berlin, Germany}\\[4mm]
{\it${}^b$Razmadze Mathematical Institute,}\\
{\it M.Aleksidze 1, 0193, Tbilisi, Georgia}

\vspace{20pt}

\end{center}
\vspace{10pt}
\vspace{40pt}

\centerline{{\bf{Abstract}}}
\vspace*{5mm}
\noindent
We discuss  timelike and spacelike minimal surfaces in $AdS_n$ using
a Pohlmeyer type reduction. 
 The differential equations for the reduced system are derived in a  
parallel treatment of both type of surfaces, with emphasis on their characteristic
differences.
In the timelike case we find a formulation corresponding to a complete gauge
fixing of the torsion. 
In the spacelike case we derive three sets of equations, related to
different parameterizations enforced by the Lorentzian signature of the metric in normal space. On the basis of these equations, we prove that there are no flat spacelike
minimal surfaces in $AdS_n,~n\geq 4$ beyond the four cusp surfaces used in the
Alday-Maldacena conjecture. Furthermore, we give a parameterization of flat
timelike minimal surfaces in $AdS_5$ in terms of two chiral fields.
\vspace{15pt}
\noindent
\end{titlepage}
\newpage
\noindent
\section{Introduction}
According to a remarkable conjecture put forward by Alday and Maldacena
\cite{am}, 
the $p$-point gluon scattering amplitude at strong coupling in ${\cal N}=4$
super Yang-Mills is related to a string worldsheet in $AdS_5$ approaching a
$p$-sided polygon spanned by the lightlike momenta of the scattering process
on the conformal boundary of $AdS_5$. In ref. \cite{am} this conjecture has
been checked for $p=4$. Furthermore, taking it for granted, the breakdown
of the BDS \cite{bds} ansatz for gluon amplitudes has been anticipated by 
estimating the behaviour of the string world surface for large $p$ \cite{am2}. To fully
establish the conjectured amplitude-string correspondence one
needs to solve the generalized Plateau problem for lightlike polygonal
boundaries. Since the related mathematical literature is mostly devoted to
spaces with positive definite metric, one is faced with a deep and delicate
problem, and despite a lot of effort \cite{jevicki,mironov,dobashi,thorn}
so far no real breakthrough has been achieved beyond $p=4$. 

The worldsheets constructed in \cite{am} for $p=4$ and generic kinematics
of the gluon momenta are all $SO(2,4)$ transforms of a highly symmetric
configuration embedded in an $AdS_3\subset AdS_5$. For this $AdS_3$ solution
the worldsheet approaches a lightlike tetragon winding alternating up and
down around the conformal boundary of $AdS_3$, the cylinder $\mathbb{R}\times
S^1$,  with each side just extending in a quarter of the cylinder. By
construction the surface is minimal. On top of this, by direct inspection,
one finds that the surface is flat, too. 

Given the high symmetry of this $AdS_3$ solution it is naturally
to ask,
whether one could find solutions in the subset of flat minimal surfaces also
for e.g. $AdS_4$ and a hexagon winding in a maximal symmetric way around
$\mathbb{R}\times S^2$ or for $AdS_5$ and an octagon winding 
around $\mathbb{R}\times S^3$. Furthermore, the surface
of ref. \cite{am} is spacelike. Although we are not aware of a rigorous
proof that all solutions with lightlike closed polygonal boundaries winding
around the conformal boundary of $AdS_5$ are spacelike, we expect
this to be valid. For this reason, in respect to the
Alday-Maldacena conjecture, we concentrate on spacelike minimal surfaces.    

But in parallel a look on timelike minimal surfaces is in order. They describe 
the dynamics of strings in real time. As emphasized in
ref.\cite{jevicki} the solution of ref. \cite{am} can be obtained from a rigid
open string rotating in a plane (in its limit of infinite 
extension) by Wick rotation of both the worldsheet time and some target space
coordinates. There are also dynamical rigid string solutions in $AdS_5$
describing a string performing two independent rotations in  the ($X^1,X^2$)
and the ($X^3,X^4$)-plane which are flat \cite{ft}. In this case a Wick
rotation of these solutions does not bring us back in an $AdS_5$. But 
nevertheless, it seems to be open, whether similar to the timelike case, 
there exist flat minimal surfaces also in the spacelike case, beyond the known
tetragon solution of ref.\cite{am}, which wind in the full $AdS_5$ and cannot
be embedded in an $AdS_3$ trivially extended to $AdS_5$.\\

By classical theorems of differential geometry the embedding of surfaces in
higher dimensional manifolds is controlled by the system of Gau{\ss},
Codazzi-Mainardi and Ricci equations. If these equations are fulfilled, the
surface is fixed up to isometries in the embedding space. An early
discussion of strings in $AdS_4$ along these lines has been given
in ref.\cite{bn}.

In the present paper we follow an equivalent procedure developed
originally for the reduction of the dynamics of the $O(N)$ sigma model
\cite{pohlmeyer} and applied to the dynamics of strings in de Sitter
and anti de Sitter spaces in \cite{dvs,gt,jevicki}. Our main focus will be
on the parallel treatment for both timelike (i.e. dynamical) and spacelike
minimal surfaces and the discussion of their characteristic differences.
Based on this, we can prove that there are no flat minimal spacelike
surfaces in $AdS_n$ beyond those constructed in \cite{am}, and can
parameterize  
all flat timelike surfaces in $AdS_5$ by two free chiral fields.
We also comment
on the reduction for arbitrary dimensions $AdS_n$.
\section{The general framework for minimal surfaces in $AdS_n$}      
Minimal surfaces with coordinates $z^{\mu}=(\sigma ,\tau )$ embedded in a
space parameterized by coordinates $X^k $ are solutions of the equation
\beq
g^{\mu\nu}\left (\nabla _{\mu}\partial _{\nu}X^k(z)~+~\partial
_{\mu}X^j\partial_{\nu}X^l~\Gamma^k_{jl}(X(z))\right )~=~0~,\label{min}
\eeq
with $\Gamma ^k_{jl}$ denoting the Christoffel symbols in the embedding space,
$g_{\mu\nu}$ the induced metric and $\nabla _{\mu}$ the induced
two-dimensional covariant derivative. This guarantees 
the vanishing of all mean curvatures, and it is also the stationarity
condition for the two-dimensional volume functional (Nambu-Goto action).
Realizing $AdS_n$ as a hyperboloid in $\mathbb{R}^{2,n-1}$
\beq
(Y^0(X))^2+(Y^{0'})^2-(Y^1)^2-\dots -(Y^{n-1})^2~=~1\label{hyperboloid}
\eeq
and choosing conformal coordinates on the surface one gets from (\ref{min})
\beq
\partial \bar{\partial}Y^N(X(z))~-~\partial Y^K\bar{\partial}Y_K \ Y^N~=~0~.
\label{eom}
\eeq 
The choice of conformal coordinates gives the additional condition
\beq
\partial Y^N\partial Y_N~=~\bar{\partial}Y^N\bar{\partial}Y_N~=~0~,
\eeq
where $\partial ,~\bar{\partial}$ are defined by $\partial =\partial
_{\sigma}+\partial_{\tau},~~\bar{\partial} =\partial 
_{\sigma}-\partial_{\tau}$ for timelike surfaces and by $\partial =\partial
_{\sigma}-i\partial_{\tau},~~ \bar{\partial} =\partial 
_{\sigma}+i\partial_{\tau}$ for spacelike surfaces.

One now extends the vectors $Y,\partial
Y,\bar{\partial}Y$ to a basis  of $\mathbb{R}^{2,n-1}$ \cite{dvs,jevicki}
\beq
\{e_N\}~=~\{Y,\partial
Y,\bar{\partial}Y,B_4,\dots ,B_{n+1}\}~.\label{basis}
\eeq  
The orthonormal vectors $B_a$ pointwise span the normal space of the surface
inside $AdS_n$.  
By eq.(\ref{hyperboloid}) $Y$ is timelike. For timelike surfaces a further
timelike  vector is parallel to the surface, hence the normal space has to be
positive definite. In contrast for spacelike surfaces the second timelike
vector has to be in the normal space. With ($a,b=4,\dots ,n+1$)
\beq
h_{ab}~=~\delta _{ab}~\mbox{or}~\eta
_{ab},~~~\mbox{for timelike or spacelike surface,}\label{h}
\eeq
we require
\beq 
(B_a,B_b)~=~h_{ab}~,~~(B_a,Y)~=~(B_a,\partial Y)~=~(B_a,\bar{\partial}Y)~=~0~.
\label{transversal}
\eeq
Moving the basis (\ref{basis}) along the surface one
gets
\beq
\partial \ e_N~=~A_N^{~~K}~e_K~,~~~~~\bar{\partial}\  e_N~=~\bar
A_N^{~~K}~e_K~. \label{evol}
\eeq
Now the strategy is to find a suitable parameterization of the 
dynamical (geometrical) degrees of freedom in the entries of the matrices
$A$, $\bar A$ and to derive differential equations for the corresponding
functions, using the  equation of motion (minimal surface condition) (\ref{eom})
and the integrability condition for eq.(\ref{evol}). Then, after solving these
differential equations, the surface has to be reconstructed by integration of
(\ref{evol}). 

Introducing
\bea
\alpha(\sigma ,\tau)&=&\log (\partial Y,\bar{\partial} Y)\label{alpha}\\
u_a(\sigma ,\tau)&=&(B_a,\partial \partial Y)~,~~~\bar u_a (\sigma
,\tau)~=~(B_a,\bar{\partial}\bar{ \partial }Y)~,\nonumber\\
A_{ab}&=&(\partial B_a,B_b)~,~~~~~~~~~\bar A_{ab}~=~(\bar{\partial }B_a,B_b)~, 
\label{u}
\eea 
and using (\ref{eom}), (\ref{transversal}) one can give eqs. (\ref{evol}) a more detailed form
\bea
\partial Y&=&~~~~~~~~~~\partial Y\nonumber\\
\partial\partial Y&=&~~~~~~\partial\alpha\partial Y~~~~~~~~~~~~~~~+~u^bB_b\nonumber\\
\partial\bar{\partial}Y&=&e^{\alpha}Y\nonumber\\
\partial B_a&=&~~~~~~~~~~~~-e^{-\alpha}~u_a\bar{\partial}Y~+~A_a^{~~b}B_b~,\label{master}
\eea
as well as the equations which one gets by the replacements
$\partial\leftrightarrow\bar{\partial}$, $u_a\rightarrow \bar u_a$, $
 A_a^{~~b}\rightarrow \bar A_a^{~~b}$.
\footnote{Note that for timelike surfaces $u$ and $\bar u$ as well as $A$ and
$\bar A$ are real. On the other side, for spacelike surfaces $u$ and $A$ are
complex, and then the bar means complex conjugation.} Indices on $u,\bar u$ and
$A,\bar A$ are raised and lowered with the normal space metric $h$, see
eq. (\ref{h}). $A$ and $\bar A$ with both indices downstairs are antisymmetric.

Then, the integrability condition $\partial\bar{\partial}e_N=\bar{\partial}
\partial e_N$ for eq. (\ref{evol}) gives
\bea
\partial\bar{\partial}\alpha-e^{-\alpha}u^b\bar
u_b-e^{\alpha}&=&0~,\label{gauss}\\
\partial \bar u_a - ~A_a^{~~b}\bar u_b&=&0~,~~~~~~~~~~ \bar{\partial}u_a -
~\bar A_a^{~~b}u_b~=~0~,\label{codazzi}\\
e^{-\alpha}\left (\bar u_au^b-u_a\bar u^b\right )&=&\partial \bar A_a^{~~b} -
\bar{\partial} A_a^{~~b}+\bar A_a^{~~c}  A_c^{~~b}- A_a^{~~c} \bar
A_c^{~~b}~.\label{ricci}
\eea

Here, a comment on the geometrical meaning of our quantities $\alpha ,~u,~A$ is
in order. Since we are using conformal coordinates,
\beq
R~=~-2\ e^{-\alpha}       \ \partial\bar{\partial}\alpha\label{R}
\eeq
is the curvature scalar on our surface. $u,~\bar u$ parameterize the
second fundamental forms 
$l^{c}_{\mu\nu}=(B^c,\partial _{\mu}\partial_{\nu}Y)$ with built in minimal surface
condition  $l^{c~\mu}_{~~\mu}=0$. Writing for timelike surfaces
$u=a+b$ and $\bar u=a-b$ one gets
\beq
l^c_{11}=l^c_{22}=\frac{1}{2}\ a^c~,~~~~l^c_{12}=l^c_{21}=\frac{1}{2}\ b^c~,
\eeq
and for spacelike surfaces with $u=a+ib$, $\bar u=a-ib$
\beq
l^c_{11}=-l^c_{22}=\frac{1}{2}\ a^c~,~~~~l^c_{12}=l^c_{21}=-\frac{1}{2}\ b^c~.
\label{ll}
\eeq
The matrices $A,~\bar A$ in (\ref{codazzi}),(\ref{ricci})
describe the torsion of the surface (for $AdS_n,~n\geq\nolinebreak
4$). Eqs.(\ref{gauss})-(\ref{ricci}) are the Gau{\ss}, Codazzi-Mainardi and
Ricci equations specialized to minimal surfaces in conformal coordinates. 
Eq.(\ref{gauss}) can be related to the Gau{\ss} equation in two ways. 
One version concerns
the relation between the difference of the scalar curvature of the surface and
the constant curvature of $AdS$ to the second fundamental forms (with zero mean
curvature) in the normal space in $AdS$ only. The other version concerns the
embedding in $\mathbb{R}^{2,n-1}$, now the big space is flat, and one has
one more second form, whose mean curvature is of course not zero.

The further analysis depends crucially on the signature of the induced metric
on the surface.
\section{Timelike minimal surfaces in $AdS_n$}     
In this case all quantities in (\ref{gauss})-(\ref{ricci}) are real and the 
metric in the normal space is positive definite, see (\ref{h}). $\partial$ and
$\bar{\partial}$ are the derivatives with respect to the chiral coordinates
$z=\frac{1}{2}(\sigma +\tau),~\bar z=\frac{1}{2}(\sigma -\tau)$.  Due to the antisymmetry of
$A$ and $\bar A$ one gets from eq.(\ref{codazzi})
\beq
\bar{\partial}(u^au_a)~=~0~,~~~~~\partial (\bar u^a\bar
u_a)~=~0~.\label{u-squared-0}
\eeq
Under  a conformal transformation $z\mapsto \zeta(z)$, $\bar z\mapsto
\bar\zeta(\bar z)$ the definitions (\ref{u}) imply:\\ $u\mapsto (\zeta
')^{-2}u,~~\bar 
  u\mapsto (\bar \zeta ')^{-2}\bar u$. This can be used to achieve within
the conformal gauge
\beq
u^au_a~=~1~=~\bar u^a\bar u_a~.\label{u=1}
\eeq
There are exceptional cases, if either both  or one out of $u^au_a$ and $\bar
u^a\bar u_a$   are zero. If both are zero, due to the positive definiteness,
$u$ and $\bar u$ are zero, which implies the vanishing of all second
fundamental forms (with respect to $AdS_n$). The surface is then (part of) an
$AdS_2\subset AdS_n$. The 
exceptional case $u^au_a~=~1$ and $\bar u^a\bar u_a=0$ will be postponed to
the  end of this section. 

For a given surface, the choice of the normal vectors $B_a$ in
(\ref{transversal}) is fixed only up to a $(z,\bar z)$-dependent $SO(n-2)$ 
transformation, which effects $u,\bar u$ and $A,\bar A$ as
\bea
u_a &\mapsto \Omega _a^{~~b}\ u_b~,~~~~~~~~~~~~~~~~~~~~~~~~~~
\bar u_a&\mapsto \Omega _a^{~~b}\ \bar u_b~,\nonumber \\
A_a^{~~b} &\mapsto\left (\Omega A\Omega ^{-1}+\partial \Omega \ \Omega
  ^{-1}\right ) _a^{~~b}~,~~~\bar  A_a^{~~b}&\mapsto \left (\Omega \bar
  A\Omega ^{-1}+\bar{\partial} \Omega \ \Omega   ^{-1}\right )
_a^{~~b}~. \label{gauge} 
\eea
We now want to use this gauge freedom to simplify
eqs.(\ref{gauss})-(\ref{ricci}). Starting with light cone gauge 
$\bar A~=~0$,  we get from (\ref{codazzi}) $\bar{\partial} u=0$. Then,
with a gauge transformation  depending only on $z$, we can bring $u_a$
to the form $u_a=\delta_{ a, n+1}$. There is no possibility to
simplify $\bar u$, beyond making use of (\ref{u=1}), and we continue with
\beq
u_a~=~(0,0,\dots ,1)~,~~~~\bar u_a~=~(\chi _4,\chi _5,\dots ,\chi _n,
\pm\sqrt{1- \chi\cdot\chi}~)~.\label{u=o}
\eeq 
Inserting all this into eq. (\ref{ricci}), we see that the field strength
on the r.h.s. no longer contains the commutator term and is given by
$-\bar{\partial}A$. Furthermore, due to the 
structure of the l.h.s. and the special form of $u$, $\bar u$ all its matrix
elements are zero, except those in the last row and column. Then in addition, 
with a $z$ dependent gauge transformation, acting only in the space orthogonal 
to $B_{n+1}$, we can also achieve zeros for all matrix
elements  of $A$, except those in the last row or column
\footnote{At this point our analysis is restricted to simple connected
  patches.  On the global level putting these $A$ elements to zero could be
  obstructed by nonzero holonomies along some cycles.}  
\beq
A_a^{~~b}~=~\left (
\begin{array}{rrrr}
0&\cdots &0&\lambda _4\\
0&\cdots&0&\lambda_5\\
&&&\cdot\\
&&&\cdot\\
0&\cdots&0&\lambda_{n}\\
-\lambda _4&\cdots &-\lambda _n&0
\end{array}\right )~,~~~\bar A_a^{~~b}~=~0~.
\eeq
Inserting this parameterization into (\ref{codazzi}) and (\ref{ricci}) one
finds 
\beq
\lambda _a~=~\pm \frac{\partial \chi_a}{\sqrt{1-
    \chi\cdot\chi}}~,~~~~~~\bar{\partial }\lambda _a~=~-e^{-\alpha}\chi _a ~.
\eeq
After this complete gauge fixing we arrive at a nonlinear coupled system
of second order differential equations for the $(n-2)$ functions $\alpha ,~\chi
_4,\dots ,\chi _n$
\bea
\partial\bar{\partial}\alpha ~\mp ~\sqrt{1-\chi
  \cdot\chi}~~e^{-\alpha} ~-e^{\alpha}&=&0~,\\
\partial\bar{\partial}\chi _b ~\pm ~\sqrt{1-\chi
  \cdot\chi}~~e^{-\alpha}~\chi _b~+~\frac{\chi\cdot \bar{\partial}\chi}{1-\chi
  \cdot\chi}~\partial \chi _b&=&0~.
\eea
These equations have a similar structure to those derived for the $O(N)$
sigma model in \cite{pohlmeyer2}.

For $AdS_3$ there are no $\chi _a$, and one ends with one equation for
$\alpha$: $\partial\bar{\partial}\alpha -2\cosh{\alpha}=0$ or
$\partial\bar{\partial}\alpha -2\sinh{\alpha}=0$, depending on whether
the signs of $u_4$ and $\bar u_4$ are equal or opposite. In
refs. \cite{dvs,jevicki} only the $\sinh $ version is discussed. 
 
For $AdS_4$ besides $\alpha$, there is only $\chi _4$. With
the parameterization $\pm \sqrt{1-\chi_4^2}=\cos \beta $ one gets \cite{dvs}
\bea
\partial\bar{\partial}\alpha -e^{-\alpha}\cos\beta -e^{\alpha}&=&0\nonumber\\
\partial\bar{\partial}\beta +e^{-\alpha}\sin\beta &=&0~.\label{AdS_4}
\eea     

We still have to comment the one exceptional case $u^au_a=1,~~\bar
u^a\bar u_a=0$, postponed above. Repeating the arguments of the generic
case, but with all $\bar u_a=0$, one further gets
$\partial\bar{\partial}\alpha - e^{\alpha}=0,~~~ u_a=\delta_{ a, n+1}$ and all
$A_a^{~~b},~\bar  A_a^{~~b}$ zero. This gives a constant curvature surface 
isometric to  $AdS_2$. But since one of the second fundamental
forms is not identically zero, the embedding in $AdS_n,~n>2$ is 
not totally geodesic.
\section{Spacelike minimal surfaces in $AdS_5$} 
Now $\partial $ and $\bar{\partial}$ are the derivatives with respect to the
surface complex coordinates 
$z=\frac{1}{2}(\sigma + \nolinebreak i\tau), ~\bar z=\frac{1}{2}(\sigma -i\tau)$ and
the 
bar on $u^c$ and $A_b^{~c}$ implies complex conjugation, too. Eq. 
(\ref{u-squared-0})
holds as in the timelike case, and by a conformal (holomorphic) transformation
$z\mapsto \zeta (z),~\bar z\mapsto \overline{\zeta (z)}$ one can achieve
 eq. (\ref{u=1})  (the exceptional case $u^au_a=0$ we discuss later).
With $u_c=a_c+ib_c$ this means  
\beq
a^c\ a_c~-b^c \ b_c~=~1~,~~~~~~~~~~a^c~b_c~=~0~.\label{ab}
\eeq
The sign of $a^c\ a_c$ and $b^c \ b_c$ is indefinite. However, in a space
with just one timelike direction, see (\ref{h}), the second equation in
 (\ref{ab}) forbids that both of these terms are negative. Therefore we
end up with three cases: $ b^c \ b_c~>~0~;~~-1~\leq ~b^c \ b_c~<~0~;~~
 b^c \ b_c~=0~.$

Unfortunately, we did not find yet a simple completely gauge fixed formulation
similar to the previous section for generic $AdS_n$. For this reason
we now consider $AdS_5$, which after all is our main focus. 

Making
use of the gauge freedom (\ref{gauge}), but now with $\Omega \in O(1,2)$,
one can give $u^c$ the following form (taking $B_4$ as the timelike vector
in the normal space and $\beta $ real)
\bea
\mbox{spacelike I}&~(b^c \ b_c~>~0)~,&~~u^c~=~\big (0,i\sinh 
\frac{\beta}{2},\cosh \frac{\beta}{2} \big )\label{I}\\
\mbox{spacelike II}&~(-1~\leq ~b^c \ b_c~<~0)~,&~~u^c~=~\big (i\sin\frac{\beta}{2},\cos
\frac{\beta}{2},0\big )\label{II}\\
\mbox{spacelike III}&~(b^c \ b_c~=0)~,&~~u^c~=~(1+i\beta,1+i\beta,1)~.\label{III}
\eea
We now discuss case spacelike I in some detail. As input in the Gau{\ss}
equation (\ref{gauss}) one gets $u^c\bar u_c= \cosh\beta$.
Inserting the $u$-parameterization (\ref{I}) into (\ref{codazzi}) one finds
$A_5^{~6}=-\frac{i}{2} \partial\beta$ and the condition $iA_4^{~5}\sinh\frac{\beta}{2}
=A_4^{~6}\cosh \frac{\beta}{2} $, which leads to the parameterization
$A_4^{~5}=\rho \cosh\frac{\beta}{2}~,~~A_4^{~6}=i\rho\sinh\frac{\beta}{2}$. 
Eq. (\ref{ricci}) then gives three more differential equations for
$\beta ,~\rho ,~\bar{\rho}$, and altogether we end up with\\[3mm]
\underline{case spacelike I} ($u^c$ from (\ref{I})):\\
\beq
A_5^{~6}=-\frac{i}{2}~ \partial\beta ~,~~~A_4^{~5}~=~\rho
\cosh\frac{\beta}{2}~,~~A_4^{~6}=i\rho\sinh\frac{\beta}{2}~.\label{AI}
\eeq
\bea
\partial\bar{\partial}\alpha ~-~e^{-\alpha}~\cosh\beta
~-~e^{\alpha}~=~0~,\label{alphaI} \\
\partial\bar{\partial}\beta
~+~(e^{-\alpha}~+~\rho\bar{\rho})~\sinh\beta~=~0~,\\ 
(\bar{\rho}\partial\beta -\rho\bar{\partial}\beta )~\sinh\frac{\beta}{2}
~+~(\partial\bar{\rho}-\bar{\partial}\rho)~\cosh\frac{\beta}{2}~=~0~, \\
(\bar{\rho}\partial\beta +\rho\bar{\partial}\beta )~\cosh\frac{\beta}{2}
~+~(\partial\bar{\rho}+\bar{\partial}\rho)~\sinh\frac{\beta}{2}~=~0~.
\eea
Similarly one gets for\newpage
\noindent 
\underline{case spacelike II} ($u^c$ from (\ref{II})):\\
\beq
A_4^{~5}~=~\frac{i}{2}~ \partial\beta ~,~~~A_4^{~6}~=~\rho
\cos\frac{\beta}{2}~,~~A_5^{~6}=i\rho\sin\frac{\beta}{2}~.\label{parA}
\eeq
\bea
\partial\bar{\partial}\alpha ~-~e^{-\alpha}~\cos\beta
~-~e^{\alpha}~=~0~,\label{alphaII} \\
\partial\bar{\partial}\beta
~+~(e^{-\alpha}~+~\rho\bar{\rho})~\sin\beta~=~0~,\label{betaII}\\ 
(\bar{\rho}\partial\beta -\rho\bar{\partial}\beta )~\sin\frac{\beta}{2}
~-~(\partial\bar{\rho}-\bar{\partial}\rho)~\cos\frac{\beta}{2}~=~0~,
\label{rhoIIa} \\
(\bar{\rho}\partial\beta +\rho\bar{\partial}\beta )~\cos\frac{\beta}{2}
~+~(\partial\bar{\rho}+\bar{\partial}\rho)~\sin\frac{\beta}{2}~=~0~.
\label{rhoIIb} 
\eea
Note that the differential equations for  case II are related to those of
case I by $\beta\mapsto i\beta$. 

To be complete, we also give\\[1mm]
\underline{case spacelike III} ($u^c$ from (\ref{III})):\\
\beq
A_4^{~5}~=~\rho~,~~~~~~~~~
 A_4^{~6}~=~-A_5^{~6}~=~i\partial\beta-\rho(1-i\beta)~,~~
\eeq
\bea
\partial\bar{\partial}\alpha -2\cosh\alpha &=&0~,\label{alphaIII}\\
\partial\bar{\partial}\beta
+(e^{-\alpha}+\rho\bar{\rho})\beta+
(\bar{\rho}\partial\beta +\rho\bar{\partial}\beta )+
(\partial\bar{\rho}+\bar{\partial}\rho)~\frac{\beta}{2}&=&0~,\\
\partial\bar{\rho}-\bar{\partial}\rho &=&0~. \label{rhoIII}
\eea

Let us add some comments. In the formulation, given in the previous
section for timelike surfaces in $AdS_5$, we
needed 
three real valued functions $\alpha ,~\chi_4 ,~\chi _5$, obeying a system of
second order differential equations. Here we have
real $\alpha ,~\beta $ and one complex $\rho $, but since the
differential equations for $\rho ,~\bar{\rho}$ are of first order only, the
overall counting of degrees of freedom matches. 

There is of course
also a description of timelike minimal surfaces in $AdS_5$, in parallel to the
treatment of this section. The resulting differential equations coincide
with those for case spacelike II up to one difference: in
eqs.(\ref{rhoIIa}),(\ref{rhoIIb}) $\rho$ has to be replaced by $-\rho $. 
But the crucial
point is that per se $\rho \bar{\rho}$ can have both signs, while it is
positive semidefinite for spacelike surfaces. This will have far reaching
consequences for the existence of flat minimal surfaces, as will be discussed
in the next sections.

For $AdS_3$ there is only one $u$, namely $u^4$ and no $\rho$. Then
eq.(\ref{ab}) means $-a_4a_4+b_4b_4=1$ and $a_4b_4=0$. This necessarily
implies $a_4=0$ and $b_4=\pm 1$, hence $u_4\bar u^4=-1$, and one is left with 
the
$\sinh$-Gordon equation for $\alpha$. In contrast to the timelike case, here
the $\cosh$ variant is excluded.   

In $AdS_4$ there is not enough freedom to realize cases spacelike I or
spacelike III, one also has not to introduce $\rho$. The equations for
$\alpha$ and $\beta$ then have the same form (\ref{AdS_4}) as in the timelike
case. 

We close with the discussion of the postponed exceptional case
$u^au_a=\nolinebreak 0$. Instead of (\ref{ab}) one has $ a^c\ a_c-b^c \ b_c=0,~~a^c~b_c=0$,
implying $b^cb_c\geq 0$. To avoid a treatment in all details, let us
concentrate on the issues relevant for the search for flat surfaces in the next
section.  The case $b^cb_c=0$ gives $u^a\bar u_a=0$ and via
(\ref{gauss}) and (\ref{R}) a spacelike surface of constant negative scalar
curvature, i.e. $\mathbb{H}^2$. The case $b^cb_c>0$ allows a parameterization 
$u^a\bar u_a=e^{\beta}$. This leads to the absence of flat solutions
of (\ref{gauss}) within the exceptional cases.
\section{Flat spacelike minimal surfaces}
On a flat surface one can always choose coordinates in which the induced
metric is $\eta _{\mu\nu}$ or $\delta_{\mu\nu}$, respectively. However,
we have already completely used up the freedom of coordinate
transformations by first starting with conformal coordinates and then using
the remaining conformal transformations to get (\ref{u=1}). Therefore, for
flat surfaces we have to allow also non constant $\alpha$ with 
$\partial\bar{\partial}\alpha =
0$, see eq.(\ref{R}). 

Let us start with $AdS_3$. Then from the $\sinh$-Gordon equation one
necessarily gets $\alpha =0$. The matrices $A_N^{~K}$ and  $\bar A_N^{~K}$
that
have to be  used in the
surface reconstruction equation (\ref{evol}) are  (the timelike case
has been discussed in \cite{jevicki}, where all entries were real)
\beq
A_N^{~K}~=~\left (
\begin{array}{rrrr}
0&1&0&0\\
0&0&0&-i\\
1&0&0&0\\
0&0&-i&0\\
\end{array}\right )~,~~~~~
\bar A_N^{~K}~=~\left (
\begin{array}{rrrr}
0&0&1&0\\
1&0&0&0\\
0&0&0& i\\
0& i&0&0\\
\end{array}\right )~.\label{A-const}
\eeq
Above we had as an alternative $u_4=\pm i$, we take here $u_4=i$. The other
choice can be generated by $B_4\mapsto -B_4$ or $\tau\mapsto -\tau $ and 
describes a surface related by a sign reversal of one of the embedding
coordinates in $\mathbb{R}^{2,2}$.
  
The solution of (\ref{evol}) is now
\beq
e_N(\sigma ,\tau )={\cal M}_N^{~K}~e_K(0,0)~,~~~~{\cal M}_N^{~K}=\left
  (\exp\big (\frac{\sigma +i\tau}{2}~A\big )~\exp \big (\frac{\sigma
    -i\tau}{2}~\bar A\big )\right )_N^{~K}~.\label{frame}
\eeq
The explicit exponentiation yields
\beq
~{\cal M}_N^{~K}~=~\left (
\begin{array}{rrrr}
C_{\s}C_{\tau}&i~\bar U_{\s, \tau} &-i~ U_{\s,\tau} &S_{\s}S_{\tau}\\
-i~U_{\s ,\tau}&C_{\s}C_{\tau}&-i~S_{\s}S_{\tau}&
  \bar U_{\s ,\tau}\\
i~\bar U_{\s
  ,\tau}&i~S_{\s}S_{\tau}&C_{\s}C_{\tau}&U_{\s ,\tau} \\ 
S_{\s}S_{\tau} &U_{\s ,\tau} & \bar U_{\s,\tau}&C_{\s}C_{\tau}\\
\end{array}\right )~,\label{calM}
\eeq
with
\beq
C_{\s}=\cosh \frac{\s}{\sqrt 2}~,~~ S_{\s}=\sinh\frac{\s}{\sqrt 2}~, ~~U_{\s
  ,\tau}= \frac{1+i}{2\sqrt 2}~\big (\sinh\frac{\s +\tau}{\sqrt 2}~+~i~\sinh
\frac{\s -\tau}{\sqrt 2}\big ) ~.
\eeq
Eq.(\ref{frame}) fully describes the evolution of our adapted frame $\{e_N\}$
along the surface in terms of an initial choice at some starting point. The
freedom in this initial choice is related to isometry transformations of the
surface as a whole. Since $Y(\sigma ,\tau)$ is our first vector in the frame,
we can read off the coordinates of the surface vector with respect to
the  $\mathbb{R}^{2,n-1}$ basis $\{e_N(0,0)\}$ from the first row of the matrix
${\cal M}$. There is however still one subtlety, due to the fact that the
second and third vector of our frame are not normalized and not orthogonal
to each other
\footnote{For a fully orthonormal choice of the $e_N$ the matrix ${\cal M}$
would be $\in SO(2,n-1)$.}. Orthonormal combinations of $\partial Y$
and $\bar{\partial Y}$ are $\frac{1}{\sqrt{2}}e^{-\alpha /2}(\partial Y+
\bar{\partial} Y)$ and $\frac{-i}{\sqrt{2}}e^{-\alpha /2}(\partial Y-
\bar{\partial} Y)$ (in these combinations a sign ambiguity, again related to a
sign reversal of an embedding coordinate has been fixed). Therefore, to get
the coordinates of $Y$ with respect to an 
orthonormal basis in  $\mathbb{R}^{2,n-1}$, one has to take
$1/\sqrt{2}$ times the sum and
$-i/\sqrt{2}$ times the difference of the second and third entry of the first
row of ${\cal M}$. 
A last point to remember is that the two timelike vectors in our frame
sit at position 1 and 4. Taking all this into account we get
\bea
Y^0~=~\cosh\frac{\s}{\sqrt 2}\ \cosh\frac{\tau}{\sqrt
  2}~,&&Y^{0'}~=~\sinh\frac{\s}{\sqrt 2}\ \sinh\frac{\tau}{\sqrt 2}
~,\nonumber\\ 
Y^1~=~ \sinh\frac{\s}{\sqrt 2}\ \cosh\frac{\tau}{\sqrt 2}~,&&Y^2~=~
\cosh\frac{\s}{\sqrt 2}\ \sinh\frac{\tau}{\sqrt 2} ~.\label{aldmald} 
\eea
which is the solution used in \cite{am,am2} for the four-point amplitude.\\  

We now turn to the search for flat spacelike minimal surfaces in $AdS_5$.
Then from (\ref{alphaI}) and (\ref{alphaIII})  we conclude that there is no
such surface of type 
spacelike I or spacelike III.  In case spacelike II, due to (\ref{alphaII}),
flatness implies $\cos\beta =-e^{2\alpha}$. As long as $\sin\beta\neq 0$ this
gives after differentiation
\beq
\partial\bar{\partial}\beta~=~\frac{4e^{2\alpha}}{\sin\beta}~\big
(1-\frac{\cos\beta ~e^{2\alpha}}{\sin ^2\beta}\big
)~\partial\alpha\bar{\partial}\alpha~.
\eeq 
Inserting it into (\ref{betaII}) one arrives at the condition 
\beq   
4e^{2\alpha}~\partial\alpha\bar{\partial}\alpha~+~\big  (
    e^{-\alpha}+\rho\bar{\rho}\big )\big (1-e^{4\alpha}\big )^2~=~0~,
\label{rho-barrho}
\eeq
which, due to $\rho\bar{\rho}\geq 0$, cannot be fulfilled. 
\footnote{As mentioned already, for timelike surfaces $\rho\bar{\rho}$ can
  have both signs, thus allowing more options.}

Therefore, the only remaining possibility is $\sin\beta=0$, i.e. $\cos\beta
=-1$ (the option $\cos\beta = 1$ is excluded by (\ref{alphaII})). For $\rho
,~\bar{\rho}$ eqs. (\ref{rhoIIa},\ref{rhoIIb}) degenerate to
$\partial\bar{\rho} +\bar{\partial}\rho =0$. 
The matrices $A_N^{~K}$ and  $\bar A_N^{~K}$ for eq. (\ref{evol}) are then
\beq
A_N^{~K}=\left (
\begin{array}{rrrrrr}
0&1&0&0&0&0\\
0&0&0&-i&0&0\\
1&0&0&0&0&0\\
0&0&-i&0&0&0\\
0&0&0&0&0&i\rho\\
0&0&0&0&-i\rho&0
\end{array}\right )~,~~~
\bar A_N^{~K}=\left (
\begin{array}{rrrrrr}
0&0&1&0&0&0\\
1&0&0&0&0&0\\
0&0&0& i&0&0\\
0& i&0&0&0&0\\
0&0&0&0&0&-i\bar{\rho}\\
0&0&0&0&i\bar{\rho}&0
\end{array}\right )~.\label{A-const5}
\eeq 
Both matrices are block diagonal. This property will be conserved under
exponentiation. As a consequence, the new degrees of freedom relative to the
$AdS_3$ case, encoded in the lower right blocks with $\rho$ and $\bar{\rho}$,
do not influence the first row of the six-dimensional analog of
(\ref{calM}). 

One can make an even stronger statement on $\rho$ and $\bar{\rho}$. Via a gauge
transformation (\ref{gauge}), acting only in the space spanned by $B_5$ and
$B_6$, one can achieve $ \rho =\bar{\rho}=0$. This can be seen in two ways.
Firstly, with $\partial\bar{\rho} +\bar{\partial}\rho =0$ one finds zero field
strength 
components related to the lower right corner of (\ref{A-const5}). Secondly
going back to (\ref{II}) one finds that, as soon as either
$\sin\frac{\beta}{2}$ or $ \cos\frac{\beta}{2}$ are zero, $u$ and $\bar u$
are parallel. We are just interested  in $\cos\beta =-1$ i.e. $\cos
\frac{\beta}{2}=0$.  Then eq.(\ref{ricci}) leads to the vanishing of all
components of the field strength tensor already from the very beginning. 
 
Altogether this proves that all flat spacelike minimal surfaces in $AdS_5$
are realized in a subspace $AdS_3$, trivially extended into $AdS_5$, and are of
type 
(\ref{aldmald}).

This statement can be extended in a straightforward manner to $AdS_n,~n>5$.
Let us sketch the set of equations one gets instead of (\ref{AI}) -
(\ref{rhoIII}). The Gau{\ss} equations (\ref{alphaI}), (\ref{alphaII}) and
(\ref{alphaIII}) 
 remain
unchanged, which again excludes flat minimal surfaces of type spacelike I and
III . For
the remaining case spacelike II, eq.(\ref{parA}) is generalized to
$A_4^{~5}=\frac{i}{2}\partial \beta ,~A_4^{~b}=\rho ^b\cos\frac{\beta}{2},~
A_5^{~b}=i\rho ^b\sin\frac{\beta}{2},~~b=6,\dots ,n+1$. There arise no
constraints on $A_a^{~b}$ if both $a,b\geq 6$.  
In eq.(\ref{betaII}) one has to make the replacement $\rho\bar{\rho}\mapsto
\sum_{b=6}^{n+1}\rho ^b\bar{\rho}^b$ and in (\ref{rhoIIa}),(\ref{rhoIIb})
$\partial \bar{\rho}\mapsto \partial\bar{\rho}_a-A_a^{~b}\bar{\rho}_b$. Then
the flatness condition necessarily leeds to $\cos\beta =-1$ and a block
diagonal structure for $A_N^{~K},~\bar A_N^{~K}$ with the $(4\times 4)$ upper
left block of $AdS_3$ structure and a $(n-3)\times (n-3)$ lower right block. 
\section{Flat timelike minimal surfaces}
The flatness condition implies $\bar\partial\partial\alpha=0$, as above.
Together with
the sinh-Gordon equation $\bar\partial\partial\alpha-2\sinh\alpha=0$ in
$AdS_3$, this allows only the vanishing solution $\alpha=0$, which leads
to the rigid infinite rotating string of \cite{jevicki}.

In $AdS_4$ one has two equations (\ref{AdS_4}). One solution is $\alpha=0$,
$\cos\beta=-1$.  It obviously
corresponds to the $AdS_3$ case extended to $AdS_4$ trivially.
For $\alpha\neq 0$, 
 similarly to the spacelike case, one finds
\beq
(1-e^{4\alpha})^2~=~-4e^{3\alpha}\partial\alpha\,\bar\partial\alpha~.
\label{alpha=}
\eeq
Since for flat surfaces $\alpha $ has a chiral decomposition 
$\alpha=\phi(z)+\bar\phi(\bar z)$, the r.h.s of eq. (\ref{alpha=})
is given as a product of chiral and antichiral fields. Calculating
$\partial\bar{\partial}$ of the logarithm, the r.h.s. is always zero,
while the l.h.s. vanishes only for constant $\phi$ or $\bar{\phi}$. 
Altogether (\ref{alpha=})  has no solution rather than $\alpha=0$. 

But starting from $AdS_5$
one can find more flat solutions. An explicit example is the
double spin solution of ref. \cite{ft}.
We follow the scheme of the previous section.  
The timelike analogs of eqs. (\ref{alphaII})-(\ref{betaII}), as mentioned
above,  are the same. 
The equation
similar to (\ref{rho-barrho}) provides 
\beq
\rho\,\bar\rho ~=~-\frac{4e^{2\alpha}\,
\partial\alpha\,\bar\partial\alpha}{(1-e^{4\alpha})^2}~-~e^{-\alpha}~.
\label{rho-barrho=}
\eeq
Instead of (\ref{rhoIIa})-(\ref{rhoIIb}) one gets
\bea
(\bar{\rho}\partial\beta +\rho\bar{\partial}\beta )~\sin\frac{\beta}{2}
~-~(\partial\bar{\rho}+\bar{\partial}\rho)~\cos\frac{\beta}{2}~=~0~,
\label{rho1} \\
(\bar{\rho}\partial\beta -\rho\bar{\partial}\beta )~\cos\frac{\beta}{2}
~+~(\partial\bar{\rho}-\bar{\partial}\rho)~\sin\frac{\beta}{2}~=~0~.
\label{rho2} 
\eea
The crucial
point is that $\rho \bar{\rho}$ can have both signs, while it is
positive semidefinite for spacelike surfaces. 

Nontrivial flat solutions imply $\cos\beta\neq \pm 1$, i.e. $\cos\frac{\beta}{2}\neq 0$ and
$\sin\frac{\beta}{2}\neq 0$,
that allow to simplify  (\ref{rho1})-(\ref{rho2})
in the form
\beq
\sin\beta\,\partial\bar\rho+\bar\rho\cos\beta\,\partial\beta=
\rho\bar\partial\beta~,~~~~~~~
\sin\beta\,\bar\partial\rho+\rho\cos\beta\,\bar\partial\beta=
\bar\rho\,\partial\beta~.\label{eq.rho-barrho}
\eeq
Due to $\cos\beta=-e^{2\alpha}$,  eqs. (\ref{rho-barrho=}) and
(\ref{eq.rho-barrho}) yield
\beq
\partial\rho=A\,\rho +B\,\rho^3~,~~~~~~~~
\bar\partial\rho=C\,\rho +\frac{D}{\rho}~,\label{rho=rho}
\eeq
where the functions $A$, $B$, $C$ and $D$ are expressed through
$\phi(z)$, $\bar\phi(\bar z)$. Then the consistency condition
for (\ref{rho=rho})  provides an algebraic (quadratic in $\rho^2$)
equation for $\rho$. Thus, the chiral and anti-chiral free fields
$\phi(z)$ and $\bar\phi(\bar z)$ ($\alpha =\phi +\bar\phi$) parameterize all
flat timelike minimal surfaces in $AdS_5$. 

\section{Characterization by invariants of minimal surfaces in 
$AdS_n,~n\geq 4$}
While the distinction between timelike and spacelike surfaces has a clear
geometrical and physical meaning, the various cases in section 4 appeared
on a rather technical level using conformal coordinates. To find a
characterization, which is both diffeomorphism invariant as well as
invariant with respect to local isometry transformations in the normal
space, we start with defining as $F=F_{z\bar z}$ the field strength
related to $A=A_z$ and $\bar A=A_{\bar z}$, i.e. the r.h.s of
eq.(\ref{ricci}). Next we introduce for $n \geq 4$ the invariant torsion 
quantity 
\beq
T~=~\frac{1}{8 \ \vert\det g\vert}~\epsilon ^{\alpha\beta}~\epsilon ^{\mu\nu}~
\mbox{tr} (F_{\alpha\beta}F_{\mu\nu})~.\label{T}
\eeq
Evaluating in conformal coordinates and using eq.(\ref{ricci}), $T$ becomes
\beq
T~=~\frac{1}{2}~e^{-2\alpha}~\mbox{tr}~F^2~=~e^{-4\alpha}\big ((\bar
u_au^a)^2~-~ (\bar u_a\bar u^a)(u_bu^b)\big )~.\label{Tconf}
\eeq 
Due to (\ref{h}) one has $T\leq 0$ for timelike surfaces, while $T$ can have
both signs for spacelike surfaces. Furthermore, for timelike surfaces
$T=0~\Rightarrow \forall F_a^{~b}=0$. In contrast, in the spacelike case such
a conclusion cannot be drawn.
 
Resolving with respect to $ \bar u_au^a $, putting into the Gau{\ss} equation
(\ref{gauss}) and using (\ref{R}), we get with $C= (\bar u_a\bar u^a)(u_bu^b)$
\beq
R~+~2~\pm ~2~e^{-2\alpha}~\sqrt{C~+~e^{4\alpha}~T}~=~0~.\label{Tgauss}
\eeq

\noindent
\underline{Exceptional cases:}\\
All exceptional cases, discussed in the previous sections, can be summarized
by $C=0$. Then from (\ref{Tconf}) $T\geq 0$. For timelike surfaces
this necessarily means $T=0$, hence $R+2=0$. The surface is then an
$AdS_2\subset AdS_n$. For the  spacelike case the option $T=0$ gives
a surface isometrically to $\mathbb{H}^2$, and for $T>0$ one can even fix
the sign ambiguity coming from (\ref{Tgauss}) and gets
$R+2+2~T^{1/2}=0$.\\[2mm]
\underline{Non-exceptional cases:}\\  
Here the choice of coordinates on the surface can be fixed completely 
such that $C=1$. Contrary to the exceptional cases, $\alpha $ no longer
drops out of (\ref{Tgauss}), and one can express $\alpha $ in terms
of invariant quantities
\beq
e^{-4\alpha}~=~\frac{(R+2)^2}{4}~-~T~.\label{covalpha}
\eeq

Altogether, now a nice picture emerges. First of all, as a spin off, we have
proven that for {\it all} minimal surfaces in $AdS_n,~n\geq 4$
\beq
\frac{(R+2)^2}{4}~-~T~ \geq 0~.\label{bound}
\eeq  
This inequality is saturated by the exceptional cases. 

For non-exceptional timelike minimal surfaces one has $(R+2)^2-4T> 0$,
which due to $T\leq 0$ induces no further subdivision.  

For non-exceptional spacelike minimal surfaces one gets
\bea
\mbox{case I}&:&~~~~0\leq T<\frac{(R+2)^2}{4}~,\nonumber\\
\mbox{case II}&:&~~~~T\leq 0~,\nonumber\\
\mbox{case III}&:&~~~~T=0~,~~~~~\mbox{not all}~~ F_a^{~b}=0~.\label{cases}
\eea 
Note that if $T=0$ in case I or II it results in $F_a^{~b}=0$, as in the
timelike case.
\section{Conclusions}
Along the lines of refs.\cite{dvs,jevicki} we have analyzed both
timelike and spacelike minimal surfaces in $AdS_n$. We went beyond these works
in two aspects. One concerns the derivation of the differential
equations for the reduced system for $n\geq 5$ and the other concerns
the parallel treatment of both timelike and spacelike surfaces. In this
analysis we pointed out crucial differences in the respective equations.
For spacelike minimal surfaces in $AdS_n,~n\geq 5$ one finds three types
of surfaces which differ among themselves in the form of their reduced
equations, too.

Based on our analysis, we  proved that there are no flat spacelike
minimal surfaces in $AdS_n$, beyond those embedded in an $AdS_3\subset AdS_n$
(where $AdS_3$ is totally geodesic in $ AdS_n$) and used for the tetragon case
of  the Alday-Maldacena conjecture. Furthermore, a parameterization of  
all flat timelike surfaces in $AdS_5$ by two free chiral fields has been
done. 

The considerations are performed in a certain patch of 
the surface. But since the resulting differential equations yield the
globally well defined four cusp solution, the statement can be made concerning
surfaces as a whole.

We stressed that there exist flat timelike minimal surfaces in $AdS_5$, which
cannot be embedded in an $AdS_3$ subspace \cite{ft}. The fact that
their double Wick rotation does not yield a spacelike surface in $AdS_5$
is no accident and finds its deeper explanation in the theorem just stated.

The subdivision for the description of spacelike minimal surfaces, first
introduced in the discussion based on conformal coordinates, finds a
characterization in terms of the scalar curvature $R$ and a quadratic torsion
invariant $T$. We also derived a universal inequality involving $R,~T$.

There remain a lot of open problems. First of all no progress towards
minimal surfaces with higher polygonal boundaries has been achieved.

In the application to the dynamics of open or closed strings the issue of
boundary conditions inside $AdS$ becomes relevant and restricts to some extent
the allowed conformal transformations on the surface as a whole.

In addition, our analysis generated various other questions already
before it comes to the issue of boundary conditions.
The reduction of the system for generic $AdS_n$ unfolds interesting
structures relevant to the most convenient choice of parameterizing functions
and gauge fixing. One can also apply a gauge invariant description 
using group valued fields instead of connections $(A,\bar A)$. 
This approach relates the $AdS$ string dynamics to gauged WZW models
\cite{Bakas:1995bm}, similarly to the $AdS\times S$ case \cite{gt}. 
Work in this direction is in progress.
\\[10mm]
{\bf Acknowledgment}\\[2mm]
We thank Chong-Sun Chu, Nadav Drukker, Jan Plefka and Donovan Young for
useful discussions. 
This work has been supported in part by Deutsche Forschungsgemeinschaft via
SFB 647. G.J. was also supported by GNSF.


\end{document}